\documentclass[11pt]{article}
    \input amssym.def
    \input amssym.tex

    \usepackage{epsfig}

    \newcommand{\sect}[1]{\setcounter{equation}{0}\section{#1}}

    \def\be{\begin{equation}}
    \def\ee{\end{equation}}
    \def\ba{\begin{eqnarray}}
    \def\ea{\end{eqnarray}}

\begin{document}
    \title{{\bf A Bestiary of Higher Dimensional Taub-NUT-AdS Spacetimes}}

    \author{ \large Adel M. Awad\thanks{email: adel@pa.uky.edu } \,\,$^{1,2}$ and
    Andrew Chamblin\thanks{email: chamblin@mit.edu} \,\,$^{1}$ \\
    \\
    $^{1}$ Center for Theoretical Physics, \\
    Massachusetts Institute of Technology, \\
    77 Massachusetts Avenue, \\
    Cambridge, MA 02139, USA. \\ 
    \\
    $^{2}$ Department of Physics and Astronomy,\\
    University of Kentucky, Lexington, KY 40506, USA. \\
    \\ Preprint MIT-CTP-3053}

    \maketitle

    \begin{abstract}
    We present a menagerie of solutions to the vacuum Einstein equations in six,
    eight and ten dimensions. These solutions describe spacetimes which
    are either locally asymptotically adS or locally asymptotically
    flat, and which have non-trivial topology. We discuss the global
    structure of these solutions, and their relevance within the
    context of M-theory.

    \end{abstract}

    \sect{Introduction: The NUTs and bolts of higher dimensional spaces}

    As is well known, every odd dimensional sphere $S^{2k+1}$ may be
    expressed, via the `Hopf fibration', as a $U(1)$ bundle over
    ${\Bbb C}{\Bbb P}^k$ \cite{topology}:
    \[
    S^{2k+1} ~{\equiv}~ U(1) ~{\longrightarrow}~ {\Bbb C}{\Bbb P}^k
    \]
    What is perhaps less well known, is that this fibration allows
    us to place a canonical Lorentz
    metric (which is non-singular and time-orientable) on any odd
    dimensional
    sphere. Explicitly, one writes the Lorentz metric $g^L$ as
    \[
    g^L= g^R - 2 {\hat V} \otimes {\hat V},
    \]
    where $g^R$ is the standard round Riemannian metric and ${\hat V}$ is
    the one-form dual
    to the unit vector field ${\bf V}$ which is tangent to the Hopf
    fibration.
    If $Z^a$, $a=1,\dots k+1$ are complex coordinates
    for $ {\Bbb R}^{2k+2} \equiv {\Bbb C}^{k+1}$ then the Hopf 
    fibration corresponds to the $SO(2) \subset SO(2k+2)$ action :
    \[
    Z^a \rightarrow \exp (it) Z^a
    \]
    and
    \[
    {\bf V}= {\partial \over {\partial t}}.
    \]
    The case $k=1$ should be familiar because it is
    encountered in the four dimensional Taub-NUT and Taub-NUT-adS solutions of
    Einstein's equations, which play a central
    role in the construction of diverse and interesting M-theory configurations.
    Indeed, Taub-NUT is central to the supergravity realization of the
    D6-brane of type IIA string theory, and Taub-NUT-adS in four dimensions
    provided the first testbed for the adS/CFT correspondence in spacetimes
    where the asymptotic structure was only locally asymptotically adS
    \cite{tnads}, \cite{hhp}. It is therefore natural to suppose that the
    higher dimensional generalizations of these spacetimes might provide us
    with a window on some interesting new corners of the M-theory moduli
    space. 

    Our construction of these higher-dimensional spacetimes is based on
    the construction of Bais and Batenberg \cite{bnb}, who found higher
    dimensional {\it Riemannian} metrics with the same topological
    structure. In \cite{gazza} the Lorentzian sections for the
    Bais-Batenberg metrics were written down\footnote{In fact,
    in \cite{gazza} it was shown that
    four dimensional Taub-NUT space {\it is its own anti-particle}, in the
    sense that there exists a diffeomorphism in the identity component
    ${\rm Diff} _0$ of the diffeomorphism group the reverses the light cones
    everywhere. Surprisingly, there is a topological obstruction to the
    existence of such a diffeo for the six dimensional Taub-NUT solution.}.
    While the general form of these solutions was constructed in \cite{page}
    \footnote{We thank M. Taylor-Robinson for pointing this out.}, the global
    structure of these solutions has not been discussed. One purpose of this
    note is to discuss the global topology of these solutions (in particular
    the issue of spin structure), and also to discuss the behaviour of curvature
    invariants.

    Following Bais and Batenberg, let
    $\{{\cal B}, g^{\cal B}, {\cal F}^{\cal B} \}$ be a $2k$-dimensional
    Einstein-K\"ahler manifold with
    K\"ahler form ${\cal F} ^{\cal B}$ which obeys the
    Dirac quantization condition, i.e., it represents an integral class
    \[
    \left[{ 1 \over { 2 \pi} } {\cal F} ^{\cal B} \right] \in H_2( {\cal
    B}; {\Bbb
    Z})
    \]
    Then ${\cal F} ^{\cal B} $ may be realized as the curvature of an $S^1$
    bundle over ${\cal B}$. Let
    \[
    e^0 = d t + A
    \]
    where $0 \le t <4 \pi$ is a coordinate on the $S^1$ fibre and $A$
    the `potential' for the field ${\cal F} ^{\cal B}$, so that
    \[
    dA= {\cal F} ^{\cal B}
    \]
    Then we will say that a $(2k+2)$-dimensional time-orientable Lorentzian
    metric
    is a {\it Taub-NUT-type} metric if it can be written in the following
    form:
    \begin{equation}
    F^{-1} (r) dr^2 + ( r^2 + N^2) g^{\cal B} - N^2 F(r) e^0 \otimes e^0
    \end{equation}
    where $F(r)$ is some function of $r$ only and $N$ is the `NUT charge'.
    Clearly, $N$ measures the size of the $S^1$ fibre, and so in general
    it measures how `squashed' the sphere is. We will often refer to
    ${\cal B}$ as the `base space', since it is the base space for the
    fibre bundle. In the usual way, a solution describes a `NUT'
    if the fixed point set of $e^0$ (i.e., the points where $F(r)$ vanishes)
    is zero dimensional, and it describes a `bolt' if the fixed point set
    is higher dimensional. Thus, if a metric of the form (1.1) solves the
    vacuum Einstein
    equations with a negative cosmological constant, then we will say that
    it is a {\it Taub-NUT-adS} (TN-adS) solution when the fixed point set of
    $e^0$ is zero dimensional, and that it is a {\it Taub-bolt-adS} (TB-adS)
    solution when the fixed point set is higher dimensional.
    Likewise, if it solves the vacuum equations
    with vanishing Ricci scalar then we will say it is a Taub-NUT (TN)
    or Taub-bolt (TB) metric accordingly.

    \section{Six Dimensional Solutions}

    \subsection{Taub-NUT-adS solution}

    For simplicity, we will mainly consider the Euclidean sections for
    these metrics. The Lorentzian sections may be obtained by analytically
    continuing the coordinate $\tau$ and also the parameter $n$
    (i.e., one replaces $n^2$ with $- N^2$). With this in mind,
    using $S^{2} ~{\times}~ S^2$ as a base space, the Taub-NUT-adS
    solution has the following form
    \begin{eqnarray}
    ds^2&=&F(r)(d\tau+2n\cos{\theta}_1 d{\phi}_1+2n\cos{\theta}_2
    d{\phi}_2)^2+F(r)^{-1}dr^2+(r^2-n^2)(d{\theta_1}^2\nonumber\\
    & &+\sin^2{\theta}_1 {d{\phi}_1}^2+{d{\theta}_2}^2+\sin^2{\theta}_2
    {d{\phi}_2}^2),
    \end{eqnarray}
    where F is given by

    \begin{equation}
    F(r)={1 \over
    3l^2(r^2-n^2)^2}\left[3r^6+(l^2-15n^2)r^4-3n^2(2l^2-15n^2)r^2-6mrl^2-3n^4(l^2-5n
    ^2)\right]
    \end{equation}
    In order for this to describe a NUT solution, it must be the case
    that $F(r = n) = 0$, so that all of the extra dimensions collapse to
    zero size at the fixed-point set of ${\partial}_{\tau}$. This can only
    happen when the mass parameter is fixed to be
    \begin{equation}
    m_n={4n^3(6n^2-l^2) \over 3l^2}
    \end{equation}
    If the mass parameter is not fixed in this way, then $F(r_b > n) = 0$
    and we will recover a bolt solution. We will say more about these
    solutions
    below.
    Fixing the mass at the value (2.4), we find that F(r) is given as
    \begin{equation}
    F(r)={(r-n)(3r^3+9nr^2+(l^2+3n^2)r+3n(l^2-5n^2)) \over 3(r+n)^2l^2}
    \end{equation}
    In order to avoid conical singularity the fiber has to close smoothly at
    $r=n$, and consequently ${\beta}F'(r = n) = 4\pi$ where 
    $\beta = {\Delta}{\tau}$
    is the period of $\tau$. This constraint then implies that
    \begin{equation}
    \beta = 12\pi n
    \end{equation}
    However, even if we avoid a coordinate singularity in this way,
    there is still a {\it curvature} singularity at the location of the
    nut, $r = n$. Indeed, we find that the Riemann curvature invariant
    diverges as $R^{ijkl}R_{ijkl} ~{\sim}~ (r^2-n^2)^{-2}$. As first discussed
    by Bais and Batenberg \cite{bnb}, we expect such a singularity
    whenever we take ${\cal B} ~{\neq}~ {\Bbb C}{\Bbb P}^k$. We will have
    more to say about such singularities later in the paper.

    \subsection{Taub-bolt-adS solution}

    In order to have a regular bolt at $r = r_b > n$, we must satisfy the
    following two conditions simultaneously:

    \noindent (i) $F(r_{b}) = 0$

    \noindent (ii) $F'(r_{b}) = {1\over 3n}$

    \noindent Condition (ii) follows from the fact that we still want to
    avoid a conical
    singularity at the bolt, together with the fact that the period of
    $\tau$ will still be $12 \pi n$. Obviously, in this case $r^2 - n^2$ will
    not
    vanish at $r = r_b$ and so the fixed point set of the Killing field
    ${\partial}_{\tau}$
    will be a four-dimensional bolt with the topology $S^2 ~{\times}~ S^2$.
    If we impose condition (i), then we find that the mass is fixed:
    \begin{equation}
    m = m_b = {-1 \over
    6l^2}[3{r_b}^5+(l^2-15n^2){r_b}^3-3n^2(2l^2-15n^2){r_b}-3n^4(l^2-5n^2)/r_b]
    \end{equation}
    From which we may deduce that
    \begin{equation}
    F'(r_b) = {5(r^2-n^2)+l^2 \over rl^2}
    \end{equation}
    If we now impose condition (ii), we recover $r_b$ as a function of
    $n$ and $l$:
    \begin{equation}
    r_{b \pm} = {1 \over 30 n}\left(l^2\pm
    \sqrt{l^4-180n^2l^2+900n^4}\right)
    \end{equation}
    To have a real value for $r_b$ the discriminant in the above equation must be
    non-negative. Adding the previous condition to the requirement that $r_b > n$
    this leads to
    \begin{equation}
    n \le \left({3-2\sqrt{2} \over 30} \right)^{1\over 2}l
    \end{equation}
    Unlike the NUT, this solution has no curvature singularities and is 
    everywhere regular.

    \subsection{Taub-NUT and Taub-Bolt solutions}

    We may recover a NUT or bolt solution of the vacuum Einstein equations with
    a vanishing cosmological constant\footnote{We think that solutions 
    with ${\cal B} \neq {\Bbb C}{\Bbb P}^k$ in the 
    limit $ l \rightarrow \infty $ were not known before. However, it is
    worth pointing out that higher dimensional locally asymptotically flat
    solutions were studied in \cite{marika}.} 
    simply by taking the limit
    $l ~{\longrightarrow}~ \infty$ in the metric (2.2), from which we
    recover the following form for $F$:
    \begin{equation}
    F(r) = \frac{r^4/3 - 2n^2r^2 - 2mr - n^4}{(r^2 - n^2)^2}
    \end{equation}
    For the NUT case the mass is
    \begin{equation}
    m_n=-{4n^3 \over 3}
    \end{equation}
    In case of bolt solution, the mass has the form
    \begin{equation}
    m_b={1 \over 2}({r_b}^3/3-2n^2r_b-n^4/r_b)
    \end{equation}
    where $r_b$ in this case is
    \begin{equation}
    r_b=3n
    \end{equation}
    Notice that the only branch from the adS solution that contributes here (i.e.
    $l \rightarrow \infty$ ) is $r_{b+}$. Again there is a curvature
    singularity for the NUT (at $r = n$), but no singularity for the bolt.

    Since this solution is Ricci flat in six dimensions,
    we may take the product of this metric with 
    five-dimensional Minkowski space in order to obtain a 
    solution of the IIA supergravity
    theory. The size of the $S^1$ fibre is then the size of the eleventh
    dimension, and therefore determines the string coupling constant.

    \subsection{Solutions with ${\cal B} = {\Bbb C}{\Bbb P}^2$}

    The following six dimensional solution generalizes
    the Taub-NUT metric found
    by Bais and Batenberg to include a negative cosmological
    constant. This metric is given by
    \begin{eqnarray}
    ds^2 &=& F(r)(d\tau+A)^2+
    F(r)^{-1}dr^2+(r^2-n^2)d{\Sigma_2}^2,
    \end{eqnarray}
    where $d{\Sigma_2}^2$ is the metric over ${\Bbb C}{\Bbb P}^{2}$ 
    which has the following form
    \begin{eqnarray}
    d{\Sigma_2}^2&=&{du^2 \over (1+u^2/6)^2}+{u^2 \over
    4(1+u^2/6)^2}(d\psi+\cos\theta d\phi)^2+{u^2 \over
    4(1+u^2/6)}(d\theta^2+\sin^2\theta d\phi^2),
    \end{eqnarray}
    also A is given by
    \begin{equation}
    A={u^2 \over 2(1+u^2/6)}(d\psi+\cos\theta d\phi)
    \end{equation}
    In this case $F(r)$ has exactly the same form as for 
    the choice ${\cal B}=S^2 \times S^2$. We will say more about that later.
    If we take the $l ~{\longrightarrow}~ \infty$ limit of (2.3), 
    in the NUT case, we recover the form for $F(r)$ discovered 
    by Bais and Batenberg (2.11) \cite{bnb}. In accordance with 
    the Bais-Batenberg prediction, this solution is free of 
    curvature singularities. 

    \section{Eight Dimensional Solutions}

    \subsection{Taub-NUT-adS solution}

    The following metric is a result of a $U(1)$ fibration over
    $S^{2} ~{\times}~ S^2 ~{\times}~ S^2$:
    \begin{eqnarray}
    ds^2 &=& F(r)(d\tau+2n\cos{\theta}_1 d{\phi}_1+2n\cos{\theta}_2
    d{\phi}_2+2n\cos{\theta}_3d{\phi}_3)^2+F(r)^{-1}dr^2+(r^2-n^2)(d{\theta_1}^2\nonumber\\
    & &+\sin^2{\theta}_1 {d{\phi}_1}^2+{d{\theta}_2}^2+\sin^2{\theta}_2
    {d{\phi}_2}^2+{d{\theta}_3}^2+\sin^2{\theta}_3 {d{\phi}_3}^2),
    \end{eqnarray}

    \noindent where F(r) has the form
    \begin{equation}
    F(r)={5r^8+(l^2-28n^2)r^6+5n^2(14n^2-l^2)r^4+5(3l^2-28n^2)r^2-10mrl^2+5n^6(l^2-7
    n^2)
    \over
    5l^2(r^2-n^2)^3}
    \end{equation}
    In order to have a NUT solution in this case, the mass parameter m must
    be
    \begin{equation}
    m=m_n={8n^5(l^2-8n^2) \over 5l^2}
    \end{equation}
    Again by fixing the mass at the above value, we find that F(r) is given as
    \begin{equation}
    F(r)={(r-n)(5r^4+20nr^3+(l^2+22n^2)r^2+(4nl^2-12n^3)r-35n^4+5l^2n^2) \over
    5(r+n)^3l^2}
    \end{equation}
    In order to avoid a conical singularity the fiber has to close smoothly at
    $r=n$,
    and consequently ${\beta}F'(r = n) = 4\pi$ where $\beta =
    {\Delta}{\tau}$
    is the period of $\tau$. This constraint then implies
    that
    \begin{equation}
    \beta = 16\pi n
    \end{equation}
    Again, even though we have managed to avoid a conical defect there is 
    still a curvature singularity at the NUT, where we find that 
    $R^{ijkl}R_{ijkl} ~{\sim}~ (r^2-n^2)^{-2}$.

    \subsection{Taub-bolt-adS solution}
    In order to have a regular bolt solution we must impose the following
    conditions simultaneously:

    \noindent (i) $F(r_{b}) = 0$

    \noindent (ii) $F'(r_{b}) = {1 \over 4n}$
    whence the mass is fixed to be
    \begin{equation}
    m=m_b={ 1 \over
    10l^2}[{r_b}^7+(l^2-28n^2){r_b}^5+5n^2(14n^2-l^2){r_b}^3+5(3l^2-28n^2)r_b+5n^6(l
    ^2-7n^2)/r_b]
    \end{equation}
    From this we have
    \begin{equation}
    F'(r_b)={7(r^2-n^2)+l^2 \over rl^2}
    \end{equation}
    To satisfy (ii), $r_b$ must be
    \begin{equation}
    r_{b \pm}={ 1\over 56 n }\left(l^2\pm \sqrt{l^4-448n^2l^2+3136n^4}\right)
    \end{equation}
    Requiring that $r_b$ is real and greater than $n$ implies that
    \begin{equation}
    n \le \left({4-\sqrt{15} \over 56} \right)^{1\over 2}l
    \end{equation}
    Again, this solution possesses no curvature singularities.

    \subsection{Taub-NUT and Taub Bolt solutions}

    Again, if we take the limit $l ~{\longrightarrow}~ \infty$ in the function
    (3.12) we will obtain a Ricci flat solution. Doing this we obtain the
    form for F:
    \begin{equation}
    F(r) = \frac{r^6 - 5n^{2}r^{4} + 15r^{2} - 10mr + 5n^6}{5(r^2 - n^2)^3}
    \end{equation}
    In this limit the NUT mass is
    \begin{equation}
    m_n={8n^5 \over 5}
    \end{equation}
    One can obtain a bolt solution in this limit by requiring $r_b>n$. The mass
    of this solution is
    \begin{equation}
    m_b={1\over 10}({r_b}^5-5n^2{r_b}^3+15r_b+5n^6/r_b),
    \end{equation}
    where
    \begin{equation}
    r_b=4n
    \end{equation}
    There is a curvature singularity at the NUT, and the bolt is everywhere
    regular. 
    Again, since this is Ricci flat we may take the product of this solution
    with three-dimensional Minkowski space to obtain a solution of
    eleven-dimensional SUGRA with vanishing four-form field strength
    (or equivalently IIA SUGRA with the dilaton VEV set by the size of the
    $S^1$ fibre). 

    \subsection{Solutions with ${\cal B} = S^{2} ~{\times}~ {\Bbb C}{\Bbb
    P}^{2}$}

    The following metric is a result of a $U(1)$ fibration over
    $S^{2} ~{\times}~ {\Bbb C}{\Bbb P}^{2}$:
    \begin{eqnarray}
    ds^2 &=& F(r)(d\tau+A)^2+
    F(r)^{-1}dr^2+(r^2-n^2)(d{\Sigma_2}^2+d{\Omega_2}^2),
    \end{eqnarray}
    where $d{\Sigma_2}^2$ is again the metric over ${\Bbb C}{\Bbb P}^{2}$ and
    $d{\Omega_2}^2$ is the metric over $S^2$
    \begin{eqnarray}
    d{\Omega_2}^2&=&d{\theta_1}^2+\sin^2{\theta_1} d{\phi_1}^2,
    \end{eqnarray}
    also A is given by
    \begin{equation}
    A=\cos{\theta_1} d\phi_1 +{u^2 \over 2(1+u^2/6)}(d\psi+\cos\theta d\phi).
    \end{equation}
    Again the form of $F(r)$ is exactly the same as for the 
    choice ${\cal B} = S^{2} ~{\times}~ S^2 ~{\times}~ S^2$. 
    As a result these solutions will share the same properties 
    of the ${\cal B} = S^{2} ~{\times}~ S^2 ~{\times}~ S^2$ solutions. 
    In this case we have also curvature singularity for both the 
    NUT and NUT$-$adS cases (Again $R^{ijkl} R_{ijkl} ~{\sim}~ (r^2-n^2)^{-2}$), 
    but the bolt solutions are everywhere regular.

    \section{Ten Dimensional Solutions}

    \subsection{Taub-NUT-adS solution}
    The following metric is a result of a $U(1)$ fibration over 
    $S^2 ~\times~ S^2 ~\times~ S^2 ~\times~ S^2$:
    \begin{eqnarray}
    ds^2 &=& F(r)(d\tau+2n\cos{\theta}_1 d{\phi}_1+2n\cos{\theta}_2
    d{\phi}_2+2n\cos{\theta}_3d{\phi}_3+2n\cos{\theta}_4d{\phi}_4)^2+
    F(r)^{-1}dr^2+(r^2-n^2)(d{\theta_1}^2\nonumber\\
    & &+\sin^2{\theta}_1 {d{\phi}_1}^2+{d{\theta}_2}^2+\sin^2{\theta}_2
    {d{\phi}_2}^2+{d{\theta}_3}^2+\sin^2{\theta}_3 {d{\phi}_3}^2+
    {d{\theta}_4}^2+\sin^2{\theta}_4 {d{\phi}_4}^2),
    \end{eqnarray}

    \noindent where F(r) has the form
    \begin{eqnarray}
    F(r)&=&{1\over
    35l^2(r^2-n^2)^4}[35r^{10}+5(l^2-45n^2)r^8+14n^2(45n^2-2l^2)r^6+
    70n^4(l^2-15n^2)r^4\nonumber\\
    &+&35n^6(45n^2-4l^2)r^2-70mrl^2+35n^8(9n^2-l^2)]
    \end{eqnarray}
    In order to have a NUT solution in this case, the mass parameter m must be
    \begin{equation}
    m=m_n={64n^7(10n^2-l^2) \over 35l^2}
    \end{equation}
    Again by fixing the mass at the above value, we find that F(r) is given as
    \begin{equation}
    F(r)={(r-n)(35r^5+175nr^4+(300n^2+5l^2)r^3+(25nl^2+100n^3)r^2+(47n^2l^2-295n^4)r
    -315n^5+35l^2n^3)
    \over 35(r+n)^4l^2}
    \end{equation}
    In order to avoid a conical singularity the fiber has to close smoothly at
    $r=n$,
    and consequently ${\beta}F'(r = n) = 4\pi$ where $\beta =
    {\Delta}{\tau}$
    is the period of $\tau$. This constraint then implies
    that
    \begin{equation}
    \beta = 20\pi n
    \end{equation}
    The curvature singularity at the NUT is signalled by the divergence
    $R^{ijkl}R_{ijkl} ~{\sim}~ (r^2-n^2)^{-2}$.

    \subsection{Taub-bolt-adS solution}

    In this case the conditions we must satisfy in order to have a regular bolt
    solution are

    \noindent (i) $F(r_{b}) = 0$

    \noindent (ii) $F'(r_{b}) = {1 \over 5n}$

    \noindent so that the mass is fixed to be
    \begin{eqnarray}
    m_b&=&{ 1 \over 70l^2}[35{r_b}^9+(5l^2-225n^2){r_b}^7+
    n^2(630n^2-28l^2){r_b}^5\nonumber\\
    &+&n^4(70l^2-1050n^2){r_b}^3+n^6(1575n^2-140l^2)r_b+n^8(315n^2-35l^2)]
    \end{eqnarray}
    From this we compute
    \begin{equation}
    F'(r_b)={9(r^2-n^2)+l^2 \over rl^2}
    \end{equation}
    To satisfy (ii), $r_b$ must be
    \begin{equation}
    r_{b \pm}={ 1\over 90 n }\left(l^2\pm \sqrt{l^4-900^2l^2+8100n^4}\right)
    \end{equation}
    As we mentioned previously conditions on $r_b$ lead to the following
    constraint
    on $n$
    \begin{equation}
    n\le \left({ 5-2\sqrt{6}\over 90}\right)^{1\over2}l
    \end{equation}

    \subsection{Taub-NUT and Taub-Bolt solutions}

    If we take the $l ~{\longrightarrow}~ \infty$ limit of (4.18) we
    recover the following form for $F(r)$:
    \begin{equation}
    F(r) = \frac{5r^8 - 28n^{2}r^6 + 70n^4 - 140n^{6}r^2 - 70mr - 35n^8}{35(r^2 -
    n^2)^4}
    \end{equation}
    The NUT mass for this solution becomes
    \begin{equation}
    m_n={-64n^7 \over 35}
    \end{equation}
    In the case of bolt solution, the mass of has the form
    \begin{equation}
    m_b={1\over 70}[5{r_b}^7 - 28n^{2}{r_b}^5 + 70n^4{r_b}^3 - 140n^{6}{r_b}-
    35n^8/r_b]
    ,
    \end{equation}
    where
    \begin{equation}
    r_b=5n
    \end{equation}
    As expected the NUT is singular (at $r = n$) and the bolt is not.
    This solution is Ricci flat in ten dimensions and so we obtain a solution
    of eleven dimensional SUGRA by multiplying with a trivial time direction
    $-dt^2$. It would be interesting to have a deeper understanding of the
    `braney' interpretation of these solutions.

    \subsection{Solutions with
    ${\cal B} = S^{2} ~{\times}~ S^2 ~{\times}~ {\Bbb C}{\Bbb P}^2$ and ${\Bbb
    C}{\Bbb P}^2 ~{\times}~ {\Bbb C}{\Bbb P}^2$}
    The following metric is a result of a $U(1)$ fibration over
    $S^{2} ~{\times}~ S^{2} ~{\times}~ {\Bbb C}{\Bbb P}^{2}$:
    \begin{eqnarray}
    ds^2 &=& F(r)(d\tau+A)^2+
    F(r)^{-1}dr^2+(r^2-n^2)(d{\Sigma_2}^2+d{\Omega_2}^2+d{{ \Omega'}_2}^2),
    \end{eqnarray}
    where $d{\Sigma_2}^2$ are again the metrics over ${\Bbb C}{\Bbb P}^{2}$,
    $d{\Omega_2}^2$ and $d{{\Omega'}_2}^2$ are the metrics over $S^2$.
    A is given by
    \begin{equation}
    A=\cos{\theta_1} d\phi_1+\cos{\theta_2} d\phi_2 +{u^2 \over
    2(1+u^2/6)}(d\psi+\cos\theta d\phi).
    \end{equation}
    The coordinates with subscript ``1'' and ``2'' are 
    the coordinates of the spheres but those without subscript 
    are the coordinates of ${\Bbb C}{\Bbb P}^2$.
    One can construct another solution using a $U(1)$ fibration over
    ${\Bbb C}{\Bbb P}^{2} ~{\times}~ {\Bbb C}{\Bbb P}^{2}$ instead:
    \begin{eqnarray}
    ds^2 &=& F(r)(d\tau+A)^2+
    F(r)^{-1}dr^2+(r^2-n^2)(d{\Sigma_2}^2+d{ {\Sigma'}_2}^2),
    \end{eqnarray}
    where $d{{\Sigma'}_2}^2$ is another metric over ${\Bbb C}{\Bbb P}^{2}$.
    A in this case is given by
    \begin{equation}
    A={u'^2 \over 2(1+u'^2/6)}(d\psi'+\cos\theta' d\phi') +{u^2 \over
    2(1+u^2/6)}(d\psi+\cos\theta d\phi).
    \end{equation}
    Again the form of $F(r)$ for both cases is exactly the same 
    as for the choice ${\cal B} = S^{2} ~{\times}~ S^2 ~{\times}~ S^2 ~{\times}~
    S^2$ 
    and the three cases share the same properties. 
    In these three cases there is a curvature singularity for both the 
    NUT and NUT-adS solutions (i.e. $R^{ijkl}R_{ijkl} ~{\sim}~ (r^2-n^2)^{-2}$), 
    but the Bolt and Bolt-adS solutions are regular everywhere.

    \section{Global structure: adS/CFT on exotic manifolds?}

    We have presented solutions of the vacuum Einstein equations with
    a negative cosmological constant, which are only locally asymptotic
    to adS space. We now briefly digress on the global structure of
    these solutions, and what we might hope to learn within the context of
    the adS/CFT correspondence.
    As is well known, the adS/CFT correspondence asserts that the
    propagators for a large N supersymmetric CFT on the boundary
    ${\partial} M$ of some locally asymptotically adS space M, are
    actually equivalent to supergravity partition functions in the bulk
    of M. In spite that most of the studied cases in the adS/CFT context were only
    anti--de--Sitter 
    spacetimes in 2, 3, 4, 5, and 7 dimensions, it has been shown that the massive 
    type IIA theory admits a wrapped-product solution of adS$_{6}$ with $S^4$
    which turns 
    out to be the near-horizon geometry of a semi-localized D$_{4}$/D$_{8}$ brane
    intersection 
    \cite{ferrara,oz,youm,justin}. Therefore, it is interesting to look for higher
    dimensional solutions--especially in six dimensions--such as Taub-NUT and
    Taub-Bolt spacetimes and study their thermodynamics in the context of the
    adS/CFT correspondence. Leaving thermodynamics of these solutions for a future
    work, here we would like to discuss the golbal structure of these solutions.

    For our solutions the boundary is generically
    a $U(1)$ fibre bundle over ${\cal B}$
    \[
    {\partial}M = S^1 ~{\longrightarrow}~ {\cal B}
    \]
    with a metric
    \begin{equation}
    ds^2_{boundary} = g^{\cal B} + {n^2 \over l^2} (d\tau + A) \otimes (d\tau +
    A)
    \end{equation}
    where $A$ is potential for the field ${\cal F}^{\cal B}$. Thus, these
    solutions provide a window of opportunity for the study of conformal
    field theories on (Euclidean) spaces with exotic topologies. Indeed,
    as in \cite{tnads}, \cite{hhp} we should be able to understand the 
    thermodynamic phase structure of a CFT on one of these spaces by working
    out the corresponding phase structure for the supergravity solutions
    in the bulk. \footnote{Skenderis \cite{kostas} has also mentioned these
    solutions as non-trivial examples to which one may apply the
    construction of a regularised energy momentum tensor.}

    Of course, one might be concerned by the appearance of naked singularities
    in the bulk of these manifolds. However, we would argue that from the point
    of view of the adS/CFT duality it may be possible to resolve these
    singularities.
    Indeed, if we invoke the criterion of Gubser \cite{steve} then these 
    singularities are `good', in the sense that a finite temperature deformation
    will yield a non-singular solution (i.e., a bolt). In fact, we would assert
    that
    our analysis leads to a rather interesting prediction: A CFT on a 
    Euclidean manifold of the form (5.51) {\it must} have some phase which
    is dual to a singular bulk supergravity solution, if 
    ${\cal B} ~{\neq}~ {\Bbb C}{\Bbb P}^k$. Put another way, changing the
    topology
    of the boundary where the CFT is defined can have dramatic consequences for
    the bulk geometry. It would be interesting to know if it is possible 
    to {\it explicitly} resolve the bulk singularity, through some mechanism
    such as that discussed in \cite{jpp}.

    It is also worth commenting on the global {\it topological} structure
    of these solutions. Recall \cite{tnads} that in four dimensions,
    Taub-bolt and Taub-bolt-adS do not admit a spin structure\footnote{Of course,
    the spacetime may still admit a $Spin^{C}$ structure}. Basically,
    this is because the bolt itself is a two-cycle with odd self-intersection
    number, i.e., the second Stiefel-Whitney class is non-vanishing on the bolt
    \cite{milnor}.

    A very similar thing happens for the solutions with
    ${\cal B} = S^{2} ~{\times}~ ... S^2~{\times}~ S^2$. For these solutions,
    it is straightforward to show that each $S^2$ factor generates an element
    of $H_2$, and that each of these elements has odd self-intersection.
    Thus, the Taub-bolt-(adS) spacetimes with this topology will not admit
    a spin structure. On the other hand, the Taub-NUT-adS solutions
    all admit spin structure.

    \subsection{The General Form of Solutions in Higher Dimensions}

    As we have noticed the forms of $F(r)$ for different choices of 
    the base space in certain dimension are exactly the same. 
    This is not a coincidence, the reason is that the differential 
    equation one can get for $F(r)$ does not depend on the 
    K\"ahler metric or the potential A. As a result the 
    function $F(r)$ depends only on the {\it dimension}, 
    $2k$ of the base space. For any $2k$ we can integrate
    the Einstein equations to obtain the general expression for $F_{2k}(r)$:
    \begin{equation}
    F_{2k}(r) = {r \over (r^2-n^2)^k} \int^r \left[{(s^2-n^2)^k \over s^2}+{2k+1
    \over l^2}{(s^2-n^2)^{k+1} \over s^2}\right] \,ds.
    \end{equation}
    Notice that the mass term here is the integration constant. As a result 
    the form of $F(r)$ for ${\cal B}={\Bbb C}{\Bbb P}^3$ and ${\cal B}={\Bbb
    C}{\Bbb P}^4$ will be the same form of $F(r)$ which we found for solutions in
    eight and ten dimensions respectively (again, these solutions should be
    regarded as the generalization of the Bais-Batenberg solutions to negative
    cosmological constant in eight and ten dimensions).

    \sect{Conclusion: Global structure of the Lorentzian sections}

    Here we have focussed on the Euclidean sections for
    the solutions. As we mentioned, we may recover the Lorentzian
    sections simply by analytically continuing $\tau$ and simultaneously
    sending $n^2$ to $-N^2$. We feel it is worth commenting on the
    causal structure of these solutions.

    First of all, since we are in Lorentzian signature, roots of
    $g_{tt} = F(r)$ correspond to horizons in the spacetime. In fact,
    a NUT (or bolt) is no longer a zero dimensional (or two dimensional)
    fixed point set, but rather it
    is a {\it chronology} horizon in the spacetime. What this means is that
    we can move from a region containing no closed timelike curves
    (the region where the coordinate $r$ is timelike), to a region containing
    closed timelike curves (the region where the compact coordinate $t$ is
    timelike) by moving across this horizon. So the Lorentzian sections of
    these manifolds generically contain closed timelike curves in the bulk.

    Furthermore, since $n^2 ~{\longrightarrow}~ -N^2$ the metric component
    in front of the base space part of the metric is now
    \[
    r^2 + N^2
    \]

    This means that the horizon region is always non-singular (the curvature
    singularities only manifest themselves on the {\it Euclidean} section)
    and so we may extend through the horizons without obstruction.
    This also means of course that the `base space' dimensions 
    never collapse to zero size,
    and we do not need to impose any regularity condition in order to avoid
    a conical singularity. This means that the mass and NUT charge are no
    longer related (as in e.g. (2.4)), but they can be freely specified.

    {\noindent \bf Acknowledgements}\\

    We would like to thank Gary Gibbons, Roberto Emparan, Al Shapere 
    and Marika Taylor-Robinson for 
    useful conversations, and especially RE, AS and GG for suggestions and
    comments 
    on a preliminary draft of this paper. AC is supported in part by funds
    provided 
    by the U.S. Department of Energy (D.O.E.) under cooperative research agreement

    DE-FC02-94ER40818. AW is supported by funds provided by ID No.
    DE-FG02-00ER45832 
    and the graduate school fellowship, University of Kentucky.

    \end{document}